\documentstyle[12pt,psfig]{article}

\input tcilatex
\begin{document}

\begin{titlepage}
\vskip 0.3cm

\centerline{\large \bf Calculation of hyperfine splitting in mesons}
\centerline{\large \bf using configuration interaction approach}

\vskip 0.7cm

\centerline{V. Lengyel, Yu. Fekete$^{\ast}$, I.
Haysak$^{\dagger}$, A. Shpenik$^{\ddagger}$}

\vskip .3cm

\centerline{\sl Uzhgorod State University,}
\centerline{\sl Department of Theoretical Physics, Voloshin str. 32,}
\centerline{\sl 88000 Uzhgorod, Ukraine}

\vskip 1.5cm

\begin{abstract}
The spin - spin mass splitting of light, heavy and mixed mesons
are described within a good accuracy in the potential model with
screened potential. We conclude that the long - distance part of
the potential cannot be pure scalar and that a vector - scalar
mixture is favoured. With the same parameters which gives correct
average mass spectrum excellent spin - spin splittings of heavy
quarkonia is obtained. The results are obtained by going beyond
usually used perturbation method, namely using configuration
interaction approach.
\end{abstract}

\vskip .3cm

\vskip 8cm

\hrule

\vskip .9cm

\noindent
\vfill $ \begin{array}{ll} ^{\ast}\mbox{{\it e-mail
address:}} &
 \mbox{kishs@univ.uzhgorod.ua}
\end{array}
$

$ \begin{array}{ll} ^{\dagger}\mbox{{\it e-mail address:}} &
   \mbox{haysak@univ.uzhgorod.ua}
\end{array}
$

$ \begin{array}{ll} ^{\ddagger}\mbox{{\it e-mail address:}} &
 \mbox{shpenik@iep.uzhgorod.ua}
\end{array}
$

\vfill
\end{titlepage}\eject
\baselineskip=14pt

\section{Introduction}

The problem of hyperfine splitting in mesons still attract wide interest. It
is widely acceptable that quark potential model gives a rather good
description of spin-average mass spectrum of hadrons , considered as
composite system of quarks \cite{LS}. However, the question of explaining
the influence of spin, namely spin-orbit ($^{\text{,,}}$fine$^{\text{,,}}$),
spin-spin ($^{\text{,,}}$hyperfine$^{\text{,,}}$) interaction, is not solved
yet. The problem of mass splitting is due to spin structure and it is
closely connected with the Lorentz - structure of the quark potential. These
effects are far from being solved yet. Quite recently QCD motivated
potential was applied to heavy quarkonia. Unfortunately authors of \cite{CO}
restrited themselves to comparing the results to a single experimental value 
$J/\Psi -\eta _c$ mass difference. No attempt was made to calculate other
quark-antiquark pairs. Spectroscopy of heavy mesons is studied also in paper 
\cite{MM}. The authors calculated only the spectra-mass of heavy mesons $B$, 
$B_S$, $D$, $D_S$ with including spin-spin interaction. In \cite{MM} the
first and second-order calculations of masses are in good agreement with the
experimental data except for the higher spin states. Besides there are
several works, in which the authors take limited number of pairs (like $u%
\overline{u}$) of particles for which splitting are calculated. Their
results are quite good, but in work \cite{BG} authors limited themselves by
considering only $\pi $, $\rho $, $K$, $K^{*}$ mesons. In analogy Buchmuller
and Tye \cite{BT} also considers only two systems, namely $c\overline{c}$
and $b\overline{b}$ though in states 1$S$ and 2$S$. They had obtaine good
results for these particular pairs. There is another paper \cite{GK}, in
which both splitting and decay properties are studied. In this paper also
good results for hyperfine splitting are obtained. But the authors calculate
only few especial mesons ($J/\Psi -\eta _c$). Also the authors at
calculations guess that in the hyperfine splitting gives contribution only
the one - gluon term. From our paper it is visible, that confinement also
plays an essential role at calculations the hyperfine effects in mesons.

The main problem of our work is to clarify some aspects of hyperfine
interaction in the framework of configuration interaction approach (CI
approximation, or CIA \cite{FBJ}).

Following many authors we assume admixture vector-scalar potential (soft
model). We consider vector and scalar parts of static potential \cite{LS}

\begin{equation}
V\left( r\right) =V_S\left( r\right) +V_V\left( r\right)  \label{f1}
\end{equation}
where

\begin{equation}
V_V=-\frac{a_S}r+\varepsilon \frac{g^2}{6\pi \mu }\left( 1-e^{-\mu r}\right)
,V_S=\left( 1-\varepsilon \right) \frac{g^2}{6\pi \mu }\left( 1-e^{-\mu
r}\right)  \label{f2}
\end{equation}
and $\varepsilon $ is the mixing constant.

The Hamiltonian can be written as

\begin{equation}
H=H_0+H_{SS}  \label{f3}
\end{equation}
where for screened potential:

\begin{equation}
H_0=\frac 1{2m}\nabla ^2-\frac{a_S}r+\varepsilon \frac{g^2}{6\pi \mu }\left(
1-e^{-\mu r}\right)  \label{f4}
\end{equation}
where $m$- is reduced mass of $q\overline{q}$- system and, $h=c=1$ units are
used. In (\ref{f3}) only $H_{SS}$ we take into account, because we calculate
hyperfine splitting in $S$-waves. Then all terms in which contain the
orbital quantum number $l$ are absent.

In the frame of Breit-Fermi approach the spin-spin interaction term is

\begin{equation}
H_{SS}=\frac 2{3m_{q_1}m_{q_2}}\overrightarrow{S_1}\overrightarrow{S_2}%
\Delta V_V  \label{f5}
\end{equation}
where $\overrightarrow{S}_{1,2}$ are the spin of the particles.

Now, we consider Shroedinger equation

\begin{equation}
\left( H_{0}+H_{SS}\right) \Psi \left( \overrightarrow{r}\right) =E\Psi
\left( \overrightarrow{r}\right)  \label{f6}
\end{equation}

Here we suggest to use CI approach, which was previously very successfully
applied in atomic physics \cite{FBJ}. The essence of this approximation is
that wave function (r) (r) is expanded in set of eigenfunctions of the
Hamiltonian $H_0$, is

\begin{equation}
\Psi \left( \overrightarrow{r}\right) =\stackunder{n}{\sum a_{n}\varphi
_{n}\left( \overrightarrow{r}\right) }  \label{f7}
\end{equation}

With substituting (\ref{f7}) in to (\ref{f6}) and using eigenvalue $E_n^0$ ,
we obtain homogeneous system of linear equations for $a_n$

\begin{equation}
a_n\left( E_m^0-E\right) =-\stackunder{n}{\sum }a_n\left\langle \varphi
_m\left| H_{SS}\right| \varphi _n\right\rangle  \label{f8}
\end{equation}
which have to be truncated for reasonable large $n$. In (\ref{f8}) is the
eigenvalue of the nonperturbative Hamiltonian:

\begin{equation}
H_{0}\varphi _{m}=E_{m}^{0}\varphi _{m}  \label{f9}
\end{equation}

It is evident, that the solution of (\ref{f8}) exists only if the
determinant, that contains of the coefficients are equal to zero. The
diagonalization of this determinant gives the values of energies we are
looking for. This is a good method to finding the eigenvalues. This
procedure goes far outside of perturbative method.

\section{Hyperfine splitting}

In this work we obtained hyperfine splitting for heavy, light and mixed
mesons using the configuration interaction approach. For example, we
calculate the hyperfine splitting in and mesons with using oscillator
potential.

Let us write down for this case Fermi-Breit equation for the two-quark
system in the form

\begin{equation}
\left( -\frac 1{2m}\Delta +Ar^2-\frac{\alpha _S}r+H_{SS}\right) \Psi \left( 
\stackrel{\rightarrow }{r}\right) =E\Psi \left( \stackrel{\rightarrow }{r}%
\right)  \label{f10}
\end{equation}
where

\begin{equation}
H_{0}=-\frac{1}{2m}\Delta +Ar^{2}  \label{f11}
\end{equation}

\begin{equation}
H_{SS}=\frac 2{3m_{q_1}m_{q_2}}\overrightarrow{S_1}\overrightarrow{S_2}%
\Delta V_V  \label{f12}
\end{equation}
and (\ref{f12}) is a additional for non-perturbatived Hamiltonian.

In this case vector and scalar part of the potential is equal accordingly:

\[
V_V=-\frac{\alpha _S}r+\varepsilon Ar^2\text{ and }V_S=\left( 1-\varepsilon
\right) Ar^2. 
\]

From one hand the oscillator potential not so bad in describing quarkonia
data, on the other hand allowing to obtain analitic basic solutions. $H_{SS}$
is the additional term, which have to be taken into account in CIA. Also one
- gluon exchange term we put in to CIA. Below for obtaining the reliable
results we shall use more realistic screened potential and make the
numerical evaluation of the appropriate matrix elements. But for a while let
us remain at using oscillator basis functions. Substituting the expanding
the wave function into (\ref{f10}) and using the condition of ortonormality
of the basic functions we obtain the algebraic system of equations for the
coefficients $a_n$.

Two remarks are to be done in connection with (\ref{f12}). First we shall
consider the contribution to spin-spin term of the potential consisting of
both one-gluon and many-gluon exchanges (\ref{f2}). Therefore

\begin{equation}
H_{SS}=\frac 2{3m_{q_1}m_{q_2}}\overrightarrow{S_1}\overrightarrow{S_2}%
\left( 4\varepsilon \pi \alpha _S\delta \left( \stackrel{\rightarrow }{r}%
\right) +6A\right)   \label{f13}
\end{equation}
$\overrightarrow{S_1}\overrightarrow{S_2}=-\frac 34$ for psevdoscalar mesons
and $\overrightarrow{S_1}\overrightarrow{S_2}=\frac 14$ for vector mesons.
In the latest calculations we shall take the interaction as a screened
potential and we shall obtain the basic functions numerically and evaluate (%
\ref{f6}) with these functions. The second remark concerns to the meaning of
parameter $\varepsilon $. According to general point of view \cite{LS} only
vector contribution is present in (\ref{f5}). In order to avoid introducing
additional parameters we shall consider the mixing parameter e to be same
for many-gluon terms approximately equal to $\varepsilon =0.5$.

If we restrict ourselves by one term in expansion (\ref{f7}) then (\ref{f8})
reduce to perturbation method. We present the wave function in (\ref{f10})
as two basic function of (\ref{f7}),

\begin{equation}
\Psi \left( \stackrel{\rightarrow }{r}\right) =a_1\varphi _1\left( \stackrel{%
\rightarrow }{r}\right) +a_2\varphi _2\left( \stackrel{\rightarrow }{r}%
\right) ,  \label{f14}
\end{equation}
immediately leads to a much better approximation. Namely one obtains for the
energies

\begin{equation}
E_{1,2}=\frac{E_1^0+E_2^0+H_{11}+H_{22}}2\pm \frac 12\sqrt{\left(
E_1^0-E_2^0+H_{11}+H_{22}\right) ^2+4H_{12}H_{21}},  \label{f15}
\end{equation}
where $H_{nm}=\left\langle \varphi _m\left| H_{SS}\right| \varphi
_n\right\rangle $.

Let us write for completness the first two terms of oscillator wave-functions

\[
\varphi _{1S}\left( r\right) =\left[ \frac{2\beta ^{3/4}}{\pi ^{1/4}}\right]
e^{-\frac{\beta r^{2}}{2}}Y_{m}^{l} 
\]

\[
\varphi _{1S}\left( r\right) =\left[ \frac{\sqrt{6}\beta ^{3/4}}{\pi ^{1/4}}%
\right] \left( 1-\frac 23\beta r^2\right) e^{-\frac{\beta r^2}2}Y_m^l 
\]
where $\beta =\sqrt{Am_q}$ . In this case the corresponding matrix elements
should be have the form

\[
H_{11}=-\frac{2\alpha _{S}\beta ^{1/2}}{\pi ^{1/4}}+\frac{\overrightarrow{%
S_{1}}\overrightarrow{S_{2}}}{m_{q}^{2}}\left\{ 4A+\frac{8\alpha _{S}\beta
^{3/2}}{3\pi ^{1/2}}\right\} 
\]

\begin{equation}
H_{22}=-\frac{5\alpha _{S}\beta ^{1/2}}{3\pi ^{1/4}}+\frac{\overrightarrow{%
S_{1}}\overrightarrow{S_{2}}}{m_{q}^{2}}\left\{ 4A+\frac{4\alpha _{S}\beta
^{3/2}}{\pi ^{1/2}}\right\}  \label{f16}
\end{equation}

\[
H_{12}=H_{21}=-\frac{2\alpha _{S}\beta ^{1/2}}{\sqrt{6}\pi ^{1/2}}+\frac{%
\overrightarrow{S_{1}}\overrightarrow{S_{2}}}{m_{q}^{2}}\left\{ \frac{4\sqrt{%
6}\alpha _{S}\beta ^{3/2}}{3\pi ^{1/2}}\right\} 
\]

Substituting (\ref{f16}) into (\ref{f15}) and using standart values of $%
E_m^0 $ for oscillator potential one can obtain the values for hyperfine
splitting which are given in the Table 1. The parameters are chosen to be $%
A=0.014$ GeV$^3$, $\varepsilon =1$, $m_q=1.5$ GeV, $\alpha _S=0.32$ since
exactly these parameters give the best results for oscillator potential in
meson masses.

\smallskip

Table1.

\smallskip

{\footnotesize 
\begin{tabular}{|cccccc|ccccc|}
\hline
\multicolumn{1}{|c|}{} & \multicolumn{1}{c|}{1} & \multicolumn{1}{c|}{2} & 
\multicolumn{1}{c|}{3} & \multicolumn{1}{c|}{4} & \multicolumn{1}{c|}{5} & 
\multicolumn{1}{|c|}{6} & \multicolumn{1}{c|}{7} & \multicolumn{1}{c|}{8} & 
\multicolumn{1}{c|}{9} & $\Delta M_{EXP}$ \\ \hline
\multicolumn{1}{|c|}{$1S$} & \multicolumn{1}{c|}{$37$} & \multicolumn{1}{c|}{%
$41$} & \multicolumn{1}{c|}{$44$} & \multicolumn{1}{c|}{$46$} & 
\multicolumn{1}{c|}{$47$} & \multicolumn{1}{|c|}{$48$} & \multicolumn{1}{c|}{%
$49$} & \multicolumn{1}{c|}{$50$} & \multicolumn{1}{c|}{$51$} & $117$ \\ 
\hline
\multicolumn{1}{|c|}{$2S$} & \multicolumn{1}{c|}{$-$} & \multicolumn{1}{c|}{$%
39$} & \multicolumn{1}{c|}{$43$} & \multicolumn{1}{c|}{$46$} & 
\multicolumn{1}{c|}{$48$} & \multicolumn{1}{|c|}{$49$} & \multicolumn{1}{c|}{%
$51$} & \multicolumn{1}{c|}{$52$} & \multicolumn{1}{c|}{$53$} & $95$ \\ 
\hline
\end{tabular}
}

\medskip\ 

Certainly using oscillator potential we obtain only didactic value.
Therefore next we turned to use more realistic potential, namely, screened
potential.The concrete form of such screened potential was previously
suggested in \cite{CJP}. We choose screened potential, because it gives
excellent description of mass-spectrum in nonrelativistic potential models
for heavy mesons. As Gerasimov pointed out \cite{SG} the QCD calculations on
lattice indicates that the spin-spin forces are rather short range. Exactly
the screened potential satisfies this condition. In this case even the basic
solutions for unperturbed Hamiltonian can not be found in analitic form. We
found these solutions numerically and evaluated the matrix elements
numerically too. Final results for hyperfine splitting for screened
potential are given in Tables 1,2. The following parameters were used in
screened potential $\frac{g^2}{6\pi }=0.224$ GeV$^2$, $\mu =0.054$ GeV. All
parameters was took from \cite{LRFCS}. Experimental values were taken from 
\cite{PDG}.

Our calculations shown that the next terms of CIA method give contribution
of order of 10\% for heavy mesons and 35\% for light mesons. Let us stress
that the first term of CIA method is in fact just a perturbation result. CIA
expansion better takes into account the interaction between particles. A
similar approach was suggested in paper \cite{NCS}, where expansion was
carried in basic function of oscillator potential. The suggested method is
of considerable interest, since the perturbation method is still used as the
practical method present \cite{MM}, \cite{BG}, \cite{CLPPS}.

\section{Discussion}

For reasons we present our data in 2 tables. We compare our results with
results obtained in the works \cite{NCS}, \cite{CLPPS}, \cite{AMB}. In the
paper \cite{NCS} good results are obtained, but the authors introduced
additional parameters r$_0$. In the paper \cite{CLPPS} hyperfine splitting
is calculated in the first account of perturbation theory. There are shown
that the first order perturbations theory gives a good description of
experimental data. The best results in paper \cite{AMB} are obtained. But
only one - gluon exchange is taken in spin-spin forces in the paper \cite
{AMB}. In all these works there is one common fault as in them are
restricted to viewing limited of number mesons.

\smallskip

Table 2.

\smallskip

{\footnotesize 
\begin{tabular}{|c|c|c|c|c|c|}
\hline
& \multicolumn{1}{c|}{
\begin{tabular}{c}
\cite{NCS} \\ 
$\Delta M_{THEOR}$ \\ 
($MeV$)
\end{tabular}
} & \multicolumn{1}{c|}{
\begin{tabular}{c}
\cite{CLPPS} \\ 
$\Delta M_{THEOR}$ \\ 
($MeV$)
\end{tabular}
} & \multicolumn{1}{c|}{
\begin{tabular}{c}
\cite{AMB} \\ 
$\Delta M_{THEOR}$ \\ 
($MeV$)
\end{tabular}
} & \multicolumn{1}{c|}{
\begin{tabular}{c}
$\text{Our results}$ \\ 
$\Delta M_{THEOR}$ \\ 
($MeV$)
\end{tabular}
} & 
\begin{tabular}{c}
$\Delta M_{EXP}$ \\ 
($MeV$)
\end{tabular}
\\ \hline
$\Delta \text{M}_{\rho -\pi }$ & \multicolumn{1}{c|}{$634$} & 
\multicolumn{1}{c|}{$550$} & \multicolumn{1}{c|}{$651$} & 
\multicolumn{1}{c|}{$923$} & $635$ \\ \hline
$\Delta \text{M}_{\rho ^{\prime }-\pi ^{\prime }}$ & $329$ & - & - & $411$ & 
$150$ \\ \hline
$\Delta \text{M}_{\varphi -\eta }$ & $217$ & - & $270$ & $580$ & $320$ \\ 
\hline
$\Delta \text{M}_{\varphi ^{\prime }-\eta ^{\prime }}$ & $135$ & - & - & $%
285 $ & - \\ \hline
$\Delta \text{M}_{K^{*}-K}$ & $405$ & $461$ & $393$ & $707$ & $398$ \\ \hline
$\Delta \text{M}_{K^{*\prime }-K^{\prime }}$ & $195$ & - & - & $336$ & $200$
\\ \hline
$\Delta \text{M}_{D^{*}-D}$ & $92$ & $147$ & $150$ & $186$ & $143$ \\ \hline
$\Delta \text{M}_{D^{*\prime }-D^{\prime }}$ & - & - & - & $112$ & - \\ 
\hline
$\Delta \text{M}_{B^{*}-B}$ & $32$ & $52$ & $58$ & $57$ & $45.9$ \\ \hline
$\Delta \text{M}_{B^{*\prime }-B^{\prime }}$ & - & - & - & $36$ & - \\ \hline
\end{tabular}
}

\medskip\ 

Table 3.

\smallskip

{\footnotesize 
\begin{tabular}{|cccccc|}
\hline
\multicolumn{1}{|c|}{} & \multicolumn{1}{c|}{
\begin{tabular}{c}
\cite{NCS} \\ 
$\Delta M_{THEOR}$ \\ 
($MeV$)
\end{tabular}
} & \multicolumn{1}{c|}{
\begin{tabular}{c}
\cite{CLPPS} \\ 
$\Delta M_{THEOR}$ \\ 
($MeV$)
\end{tabular}
} & \multicolumn{1}{c|}{
\begin{tabular}{c}
\cite{AMB} \\ 
$\Delta M_{THEOR}$ \\ 
($MeV$)
\end{tabular}
} & \multicolumn{1}{c|}{
\begin{tabular}{c}
$\text{Our results}$ \\ 
$\Delta M_{THEOR}$ \\ 
($MeV$)
\end{tabular}
} & 
\begin{tabular}{c}
$\Delta M_{EXP}$ \\ 
($MeV$)
\end{tabular}
\\ \hline
\multicolumn{1}{|c|}{$\Delta \text{M}_{D_S^{\text{*}}-D_S}$} & 
\multicolumn{1}{c|}{$87$} & \multicolumn{1}{c|}{$190$} & \multicolumn{1}{c|}{%
$128$} & \multicolumn{1}{c|}{$163$} & $144$ \\ \hline
\multicolumn{1}{|c|}{$\Delta \text{M}_{D_S^{\text{*}\prime }-D_S^{\prime }}$}
& \multicolumn{1}{c|}{-} & \multicolumn{1}{c|}{-} & \multicolumn{1}{c|}{-} & 
\multicolumn{1}{c|}{$100$} & - \\ \hline
\multicolumn{1}{|c|}{$\Delta \text{M}_{B_S^{\text{*}}-B_S}$} & 
\multicolumn{1}{c|}{-} & \multicolumn{1}{c|}{-} & \multicolumn{1}{c|}{-} & 
\multicolumn{1}{c|}{$50$} & $47$ \\ \hline
\multicolumn{1}{|c|}{$\Delta \text{M}_{B_S^{\text{*}\prime }-B_S^{\prime }}$}
& \multicolumn{1}{c|}{-} & \multicolumn{1}{c|}{-} & \multicolumn{1}{c|}{-} & 
\multicolumn{1}{c|}{$33$} & - \\ \hline
\multicolumn{1}{|c|}{$\Delta \text{M}_{B_c^{\text{*}}-B_c}$} & 
\multicolumn{1}{c|}{-} & \multicolumn{1}{c|}{-} & \multicolumn{1}{c|}{-} & 
\multicolumn{1}{c|}{$49$} & - \\ \hline
\multicolumn{1}{|c|}{$\Delta \text{M}_{B_c^{\text{*}\prime }-B_c^{\prime }}$}
& \multicolumn{1}{c|}{-} & \multicolumn{1}{c|}{-} & \multicolumn{1}{c|}{-} & 
\multicolumn{1}{c|}{$31$} & - \\ \hline
\multicolumn{1}{|c|}{$\Delta \text{M}_{\gamma -\eta _b}$} & 
\multicolumn{1}{c|}{$31$} & \multicolumn{1}{c|}{$39$} & \multicolumn{1}{c|}{$%
82$} & \multicolumn{1}{c|}{$46$} & - \\ \hline
\multicolumn{1}{|c|}{$\Delta \text{M}_{\gamma ^{\prime }-\eta _b^{\prime }}$}
& \multicolumn{1}{c|}{$9$} & \multicolumn{1}{c|}{-} & \multicolumn{1}{c|}{-}
& \multicolumn{1}{c|}{$26$} & - \\ \hline
\multicolumn{1}{|c|}{$\Delta \text{M}_{\text{J/}\Psi -\eta _c}$} & 
\multicolumn{1}{c|}{$65$} & \multicolumn{1}{c|}{$100$} & \multicolumn{1}{c|}{%
$112$} & \multicolumn{1}{c|}{$110$} & $117$ \\ \hline
\multicolumn{1}{|c|}{$\Delta \text{M}_{\Psi -}\eta _c^{\prime }$} & 
\multicolumn{1}{c|}{$32$} & \multicolumn{1}{c|}{$54$} & \multicolumn{1}{c|}{-
} & \multicolumn{1}{c|}{$67$} & $95$ \\ \hline
\end{tabular}
}

\medskip\ 

In Table 2 we show final data for light-quark systems, which has mainly
relativistic character and compare them with other data.The calculation are
carried out by using (\ref{f1}-\ref{f9}). In our quazirelativistic approach
we should take into account in Hamiltonian also the term of order p$^4$ i.
In paper \cite{AMB} it was shown that the spectrum of Hamiltonian in
nonrelativistic potential models, and spectrum of relativistic Hamiltonian
for bound state and for first radial excited state are equivalent. On the
other hand Lucha and Shoeberl \cite{LS} has shown that in some cases this
relativistic kinematik term change the mass-spectrum drustically. But since
we consider mass-difference therefore all relativistic effects must be
cancel. In other papers authors introduce new additional parameters and
obtain good description of hyperfine splitting. In paper \cite{EGF} Faustov
et al. also had made similar calculations with using quazipotential. But
evidently potential was choosen unsuccessfully, as $\varepsilon $ turned out
to be -0.9. Evidently this negative value of devoids it is clear physical
sense, as a mixed parameters. Therefore this approach lost very much in
heuristic understanding of obtained results.

In Table 3 we present the results of hyperfine splitting calculation in
heavy-quark systems. Namely exactly for these systems our Breit-Fermi
approach must be true with maximum extent. The obtained results for the
hyperfine splittings of the $S$-wave states agree with measured splittings.
As it is see most of our results have mainly predective character. In the
Table 3 are shown that for 2$S$-states we obtain somewhat worse results for
hyperfine splitting than in 1$S$-states. This can be the result of a mixing
of $S$ and $D$ waves. In 2$S$ state, as it is shown in work \cite{HLM}, the
mixing can give contribution of 10\%, while in case of 1$S$-states the
mixing correction is only of 1\%. In other words we believe that taking into
account the mixture of $S$ and $D$ waves, would considerable in proove our
results.

We suggest that the potential consist of a sum of vector and scalar parts.
This idea of scalar-vector mixing was discussed in \cite{LF}, \cite{DC1}-%
\cite{DC3}. The authors of these papers also came to the conclussion that
its mixing parameter must be different from zero. Franzinis \cite{LF} show,
that VCONF must be tottaly scalar, while VOGE must be tottaly vector, but
nevertheless they cannot give an adeqyate description of data, concerning
the fine splitting. In series of papers \cite{DC1}-\cite{DC3} Deoghuria and
Chakrabarty had choosen the confining potential in the form

We suggest that the potential consist of a sum of vector and scalar parts.
This idea of scalar-vector mixing was discussed in \cite{LF}, \cite{DC1}-%
\cite{DC3}. The authors of these papers also came to the conclussion that
its mixing parameter must be different from zero. Franzinis \cite{LF} show,
that $V_{CONF}$ must be tottaly scalar, while $V_{OGE}$ must be tottaly
vector, but nevertheless they cannot give an adeqyate description of data,
concerning the fine splitting. In series of papers \cite{DC1}-\cite{DC3}
Deoghuria and Chakrabarty had choosen the confining potential in the form

\[
V=\varepsilon V_{OGE}+\left( 1-\varepsilon \right) V_{CONF} 
\]
and found $\varepsilon =0.2$, treating as an adjustable parameter. In this
paper we have used the same approach for CJP-type potential and found $%
\varepsilon =0.5$. With this value of we described the hyperfine splitting
of all mesons from heave to light ones.


\newpage

\begin{thebibliography}{99}
\bibitem{LS}  W. Lucha, F. Schoberl, Effective potential models for hadrons
HEPHY-UB 62P1/95 UW THPH 1995-16.

\bibitem{CO}  Yu-Qi Chen, R. Oakes Phys. Rev.{\bf \ D}, Vol 53, ${}$\#9,
(1996).

\bibitem{MM}  T. Matsuki, T. Morii Phys. Rev. {\bf D}, Vol 56, ${}$\#9,
(1997).

\bibitem{BG}  L. Burakovsky, T. Goldman Phys. Rev.{\bf \ D}, Vol 57, ${}$%
\#5, (1998).

\bibitem{BT}  W. Buchmuller, S.-H. H. Tye Phys. Rev. {\bf D}, Vol 24, ${}$%
\#1, (1981).

\bibitem{GK}  S. Godfray, R. Kokoski Phys. Rev. {\bf D}, Vol 43, ${}$\#5,
(1991).

\bibitem{FBJ}  C. Froese, T. Brige, P. Johnsson, Computational atomic
structure. An MCHF approach, Inst. Phys. Publishing. Bristol, Phyladelphia
(1997).

\bibitem{CJP}  Z.Chikovani, L. Jenkovszky, F. Paccanoni, Mod. Phys. Lett. 6%
{\bf A}, 1401 (1991).

\bibitem{SG}  S.Gerasimov, Proceedings of the International Conference
''Hadron Structure-98'', Stara Lesna, Slovak Republic, 7-13 September 1998,
p.179.

\bibitem{LRFCS}  V. Lengyel, V. Rubish, Yu. Fekete, S. Chalupka, M. Salak,
Condensed Matter Physics, vol.1, ${}$\#3(15), 575 (1998).

\bibitem{PDG}  Part. Data Group, Phys. Rev. {\bf D},1 (1996).

\bibitem{NCS}  I. M. Narodetskii, R. Ceuleneer, C. Semay J. Phys. {\bf G}:
Nucl. Part. Phys.18, 1901 (1992).

\bibitem{CLPPS}  S. Chalupka, V. Lengyel, P. Petreczky, F. Paccanoni and M.
Salak, Nuovo Cimento 107{\bf A}, 1557 (1994).

\bibitem{AMB}  A. M. Badalyan J. Nucl. Phys. Vol 46, {\#}4(10), (1987).

\bibitem{EGF}  D. Ebert, V. O. Galkin, R. N. Faustov Phys. Rev. {\bf D}, Vol
57, ${}$\#9, (1998).

\bibitem{HLM}  I. Haysak, V. Lengyel, V. Morokhovych, Proceedings of the
International Conference $^{\text{,,}}$Small Triangle Meeting$^{\text{,,}}$
Kosice Slovak Republic, 8-9 September 1999, p.113-119.

\bibitem{LF}  J. Lee-Franzini, P. Franzini, Frascati-preprint,
LNF-93/064(P), Italy, 1993.

\bibitem{DC1}  S. Deoghuria, S. Chakrabarty, J. Phys. {\bf G}: Nucl. Part.
Phys.15, 1231 (1989).

\bibitem{DC2}  S. Deoghuria, S. Chakrabarty, J. Phys. {\bf G}: Nucl. Part.
Phys. 16, 185 (1990).

\bibitem{DC3}  S. Deoghuria, S. Chakrabarty, J. Phys. {\bf G}: Nucl. Part.
Phys.16, 1825 (1990).
\end{thebibliography}
\end{document}